\documentclass[12pt,epsf]{article}
\usepackage{graphicx, amsmath}

\begin{document}

\sloppy
\begin{flushright}{SIT-HEP/TM-37}
\end{flushright}
\vskip 1.5 truecm
\centerline{\large{\bf NO Curvatons or Hybrid Quintessential Inflation}} 
\vskip .75 truecm
\centerline{\bf Tomohiro Matsuda\footnote{matsuda@sit.ac.jp}}
\vskip .4 truecm
\centerline {\it Laboratory of Physics, Saitama Institute of Technology,}
\centerline {\it Fusaiji, Okabe-machi, Saitama 369-0293, 
Japan}
\vskip 1. truecm
\makeatletter
\@addtoreset{equation}{section}
\def\theequation{\thesection.\arabic{equation}}
\makeatother
\vskip 1. truecm

\begin{abstract}
\hspace*{\parindent}
We consider a curvaton scenario in which the late-time domination and
 the generation of the curvature perturbation is achieved by a
 non-oscillatory (NO) curvaton potential. 
Instead of considering the conventional curvaton oscillation, we
 consider ``weak trapping'' after preheating, which modifies the
 evolution of the curvaton density after preheating.  
The primordial isocurvature perturbation related to the curvaton is once
 converted into the fluctuation of the number density of the preheat
 field through inhomogeneous preheating.  
Then the evolution of the curvatons and the preheat field is controlled
 by the preheat-field number density. 
The density of these fields decreases slightly slower than the standard
 matter density which suggests that these fields will grow with time.
Finally, the preheat field decays to reheat the Universe leaving behind
the curvature perturbation. 
In our scenario the task of the standard curvaton is not executed solely
 by the curvaton itself but is partially shared with the preheat
 field. 
NO curvatons can be considered as the hybrid version of the
 quintessential inflationary model.   
\end{abstract}

\newpage
\section{Introduction}
\hspace*{\parindent}
Recent observation of the cosmic microwave background radiation (CMBR)
anisotropy confirms that the present structure of the Universe should
have originated from the primordial density perturbation. 
The traditional explanation for this primordial density perturbation is
that during inflationary expansion the perturbation related to the light
field (inflaton) generates the curvature perturbation during the
inflation. 
The traditional inflationary scenario, where the primordial curvature
perturbation is generated solely by the inflaton field, is very simple
but inevitably puts extra conditions on the inflationary mechanism and
the form of the inflaton potential.

Recently, alternatives to the traditional inflationary scenario
have been suggested by many authors\cite{curvaton1, matsuda_curvaton,
topological_curv, alternate, Inho_Reh_Dvali, alternate2, SSB-inst,
inst-curv}. 
The common thread of all these alternatives is that the isocurvature
perturbation
related to a light field, other than the inflaton, is converted into the
curvature perturbation after (or at the end of) inflation.
Since the seed perturbation is generated during primordial inflation,
the spatial length scale of the resultant curvature perturbation is very
large compared with the correlation length at the time when the
curvature perturbation is generated.
 There are many scenarios related to such conversion mechanisms that
 characterize the concepts behind these alternative models.
 
Among these alternatives, we will consider a mixed scenario of the
curvatons\cite{curvaton1, matsuda_curvaton} and the inhomogeneous
preheating\cite{SSB-inst, inst-curv}.
In the standard curvaton scenario the curvaton energy density is
supposed to be negligible just after inflation and the spacetime is
unperturbed at that time.  
The curvaton energy density becomes significant during some era of the
radiation domination or the era when the kinetic energy of the inflaton
dominates the Universe \cite{Quin-infla-Curvaton}.
In the case where radiation dominates the Universe, the radiation energy
density will evolve as  $\rho_{tot}\propto a^{-4}$ while the curvaton
energy density is (1) almost constant before its oscillation
($\rho_{curv}\propto a^0$), and (2) evolves as 
$\rho_{curv}\propto a^{-3}$ after it starts to oscillate.
Therefore, the ratio of the curvaton energy density to the background
radiation will grow with time, and finally the curvaton will become the
dominant component of the Universe. 
The curvaton could generate the curvature perturbation
if the curvaton oscillation lasts long time before the
decay.\footnote{There 
is a possible scenario (``heavy curvatons'')  
in which curvatons might become heavy and start to dominate the Universe 
just after inflation\cite{matsuda_curvaton}.} 
The same thing will happen if the inflaton does not decay but continues
to oscillate with a quartic potential. 
During this oscillationary period the inflaton energy density will
evolve as $\rho_{osc}\propto a^{-4}$ \cite{Toward_Kof}. 
Moreover, if the inflaton potential is non-oscillatory, the kinetic
energy of the inflaton will evolve as $a^{-6} $\cite{Quin-infla}.
In this case the domination by the curvatons will occur much earlier and
easier than in the standard curvaton scenario
\cite{Quin-infla-Curvaton}. 

Perhaps the mixed scenario that we will describe in this paper will be
categorized as the curvaton scenario in the sense that the field
accompanied by the isocurvature perturbation is 
(1) initially negligible but after a non-oscillatory period it (and the
accompanying preheat field) starts to dominate the energy density of the
Universe, and (2) the preheat field decays to reheat the Universe
generating the curvature perturbation.
Note that unlike the standard curvatons the light field (NO curvaton)
connected to the primordial isocurvature perturbation neither oscillates
nor decays to reheat the Universe. 
The isocurvature perturbation of the NO curvaton is once converted into
the fluctuation of the preheat-field number density. 
In this sense the NO curvaton seeds the generation of the curvature
perturbation that is induced by the accompanying preheat
field.\footnote{A  
similar scenario has been discussed in Ref.\cite{topological_curv},
where the nucleation rate of a decaying cosmological defect is supposed
to be biased by a light field. 
As in the NO curvatons, the light field that biases the nucleation of
the cosmological defects never dominates the energy density of the
Universe. }
The curvaton has the NO potential but weakly trapped near the ESP
because of the effective confining potential that is induced by the
preheat field.
The trapping mechanism makes the evolution of the curvaton (and the
preheat-field) energy density completely different from the conventional
scenario.  
Their evolution is determined by the force-balance equation that is
written by the NO potential and the number density of the preheat
field. 
We will show that their energy densities evolve as $a^{-3+\epsilon}$
($\epsilon =3/(n+1)$) with the constant ratio
$\rho_{preheat}/\rho_{curv}=n$. 
The mass of the preheat field grows during the evolution until it
finally reaches the critical value where the preheat field decays to
reheat the Universe. 
After the decay of the preheat field, the NO curvatons rapidly roll down
the NO potential. 
In this sense the NO curvatons could be seen as the hybrid version of
the quintessential inflation model \cite{Quin-infla}.
However, we do not stick to the idea that the NO curvaton also plays the
role of the quintessence, even though the idea seems rather
fascinating. 
The conditions for quintessential inflation are not necessary for the NO
curvatons, and vice versa.
In this paper we will focus our attention on the NO curvatons.

\section{NO curvatons}
First we briefly review the basic idea of inhomogeneous preheating.
The preheating mechanism that is presented here is basically given by
 Kofman et al \cite{Toward_Kof}.
For simplicity, we consider real scalar fields $\phi_I$, $\sigma$ and
$\chi$.
Here $\phi_I$ is the inflaton that has a chaotic-type potential. 
$\sigma$ is the light field accompanied by the primordial perturbation
and biases the production of the preheat field $\chi$.
$\sigma$ is the ``curvaton'' in our scenario, although it does not play
all the roles played by the standard curvatons.
The important roles that characterizes the curvatons are
(1) isocurvature perturbation, (2) late-time domination and (3)
reheating by the decay.
In our scenario of the NO curvatons, the first characteristic is related
to the ``curvaton'' $\sigma$, but the second and the third are connected
 to the preheat field $\chi$.  
Therefore, although the light field $\sigma$ is called ``curvaton'' in
this paper,  it does not play all the roles that have been played by the
conventional curvatons as described in earlier work.
We hope this peculiar feature of the NO curvaton does not become a
further confusion to our readers.  
With these scalar fields we will consider a model that contains the
inflation sector 
\begin{equation}
{\cal L}_I=\frac{1}{2}\partial_\mu \phi_I \partial^\mu \phi_I + 
\frac{1}{2}\partial_\mu \sigma \partial^\mu \sigma+
\frac{1}{2}\partial_\mu \chi \partial^\mu \chi 
-\frac{g^2}{2}(\phi_I^2 + \sigma^2)\chi^2
-V(\phi_I)-V(\sigma),
\end{equation}
where the cosmological constant is disregarded.
Since we are considering the chaotic-type inflation, the inflaton
potential for large $\phi_I$ in this case is given by the generic form
\begin{equation}
V(\phi_I) = \frac{\lambda_I|\phi_I|^{n_I}}{M_I^{n_I-4}},
\end{equation}
while for the curvaton potential we consider the potential
\begin{equation}
V(\sigma) = \frac{M^{n+4}}{\sigma^n}+ \frac{1}{2}m^2 \sigma^2,
\end{equation}
which has a run-away behavior if $m=0$.
Here we assume that $m$ is very small (or $m=0$) so that there is no
sensible $\sigma$-oscillation that modifies our scenario.
A similar potential could be seen for the repulsive force between
branes \cite{alternate, inst-curv}. 
 
We will consider the generation of the preheat field $\chi$ when
the inflaton $\phi_I$ approaches the origin just after chaotic inflation.
Since the mass of the preheat field ($m_\chi^2 = g^2 (\phi_I^2+\sigma^2)$)
depends both on $\phi_I$ and
$\sigma$, the generation of the preheat field will be biased if there is
 isocurvature perturbation related to $\sigma$.
Let us describe the basic idea of the preheating. 
Just after the inflation, the mass of the preheat field is very large
but will become very small when the inflaton approaches the
origin. 
Therefore, the adiabatic condition is violated if
\begin{equation}
|\dot{m}_\chi|/m_\chi^2 \sim g|\dot{\phi}_I|/g^2 (\phi_I^2+\sigma^2) >1
\end{equation}
and nonadiabatic particle production occurs.
 If the inflaton is the chaotic type, the nonadiabatic region is very
 narrow compared with the initial amplitude, and thus efficient particle
 production occurs within a very short period of time when the inflaton
 passes by the ESP.
 In this case one can neglect the expansion of the Universe during the
 particle production.
 The number density of the preheat field at the first scattering is
 calculated in Ref.\cite{Toward_Kof} and given by
\begin{equation}
\label{preh_n}
n_\chi = \frac{(gv)^{3/2}}{(2\pi)^3}\exp\left(-\frac{\pi m_\chi^2}{gv}
\right).
\end{equation}
Here $v$ is the absolute value of the inflaton velocity
at the ESP. 
We would like to suppose that the coupling constant is large ($g\sim
1$).
If the decay of the preheat field is negligible during this period, the
preheat field induces the confining effective potential for both
$\phi_I$ and $\sigma$, which leads to the 
trapping of these fields at (or near) the ESP \cite{beauty_is}.
However, as we are considering the unusual NO potential for the curvaton
$\sigma$, the trapping of the NO curvaton cannot be explained by the
standard trapping scenario.
We will discuss this exceptional trapping mechanism later in this
section.  

Before discussing the trapping of the NO curvatons, we will describe the
origin of the fluctuation that appears for the number density of the
preheat field. 
According to Eq.(\ref{preh_n}), the production is efficient if
$m_{\chi}^2 < v$.
Assuming that the potential energy during inflation is efficiently
converted into the kinetic energy of the inflaton, the velocity of the
inflaton becomes $v \simeq H_I M_p$. 
 Then the condition for the efficient production puts an upper bound on
 the initial value of $\sigma$,
\begin{equation}
\sigma < \sqrt{H_I M_p}.
\end{equation}
If the value of $M$ is very small compared with the inflationary scale,
the $\sigma$-potential is flat except for the very narrow region near the
origin. 
Therefore, we will assume that the curvaton $\sigma$ experiences a flat
potential during inflation.
The flatness condition for the curvaton is $m_\sigma < H_I$ during
inflation, which becomes 
\begin{equation}
M\left(\frac{M}{H_I}\right)^{\frac{2}{n+2}}
<\sigma.
\end{equation}
Remember that we are considering the case in which the isocurvature
perturbation related to $\sigma$  plays an important role in biasing the
production of the preheat field.
Therefore, we will consider the initial value of $\sigma$ 
\begin{equation}
M\left(\frac{M}{H_I}\right)^{\frac{2}{n+2}}
<\sigma < \sqrt{H_I M_p}
\end{equation}
so that $\sigma$ appears with the primordial isocurvature perturbation
and also does not disturb the generation of the preheat field $\chi$.
Considering the biasing effect related to the primordial perturbations
$\delta \sigma$, we obtained the fluctuation of the number
density\cite{inst-curv}
\begin{equation}
\label{num_flu}
\frac{\delta n_\chi}{n_\chi}\simeq\frac{2\pi g \sigma \delta \sigma}{v}.
\end{equation}
After the first scattering, the scatterings occur successively where the
number densities are given by 
\begin{equation}
n_k^{j+1} =  b_k^j n_k^j.
\end{equation}
Here the coefficient $b$ is given by
\begin{equation}
b_k^j =1+2e^{-\pi \mu^2}-2\sin \theta^j e^{-\pi \mu^2/2}\sqrt{
1+e^{-\pi \mu^2}},
\end{equation}
where $\mu^2=(k^2+m_\chi^2)/gv$.
Here $\theta^j$ is a relative phase.
It is possible to consider $\delta b_k^j$, but they are not important
in the present scenario because the light field $\sigma$ is 
trapped near the ESP by the strong confining potential just after the
first scattering. 
Then the small $\sigma$ makes the fluctuations appearing in $\delta b$ 
completely negligible compared with the one generated at the first
scattering. 
After successive scatterings the kinetic energy of the inflaton is
transferred into the excitations of the preheat field, which leads to
the number density \cite{beauty_is} 
\begin{equation}
n_\chi \sim v^{3/2}g^{-1/2}
\end{equation}
with the perturbation given by Eq.(\ref{num_flu}).
According to the observation, the spectrum of the curvature perturbation
on a cosmological scale is ${\cal P}_\zeta^{1/2}\simeq 2\times 10^{-5}$.
As in the standard curvaton scenario, the preheat field decays leaving
behind the curvature perturbation
\begin{equation}
\zeta \simeq \alpha r \frac{\delta \rho_\chi}{\rho_\chi},
\end{equation}
where $r$ is the ratio between $\rho_\chi$ and the total energy density
of the Universe.
The value of $\alpha$ is $\alpha=1/3$ in the standard
scenario, while in our scenario it is slightly shifted.

Now we will consider the $\sigma$-trapping.
The idea of ``trapping'' that is used in this paper is rather different
from the original idea that has been used to discuss
brane-trapping\cite{beauty_is}.\footnote{It will be very interesting to
compare our scenario with the scenario of ``Trapped quintessential
inflation'' \cite{dimo_trapped}. Although the mechanism for generating
the curvature perturbation is completely different, our present scenario
could be seen as a hybrid version of the model discussed in
Ref.\cite{dimo_trapped}. Of course we can accomodate the ``trapped
quintessential inflation'' itself by simply introducing an extra light
field that induces inhomogeneous preheating at the trapping. 
This is not the hybrid version of the original scenario.
In this case the extra light field may have a conventional (non-NO) flat
potential.} 
One might think that the NO potential is an effective potential so that
it must be regularized near the origin by an explicit cut-off. 
If so, $\sigma$ will initially be trapped at the ESP and then start to
roll down the NO potential when $n_\chi$ is diluted below some critical
value.
This speculation is conceivable, but even if there is an explicit
cut-off we can use the effective potential as far as we are considering
the late-time evolution of the curvatons.
Therefore, for the late-time evolution the effective minimum of the
$\sigma$-potential is obtained from the force-balance equation
\begin{equation}
g n_\chi(t) -\frac{n M^{n+4}}{\sigma^{n+1}}=0,
\end{equation}
which leads to the expectation value of $\sigma(t)$,
\begin{equation}
\label{rela_si}
\sigma(t) =M\left(\frac{n M^{3}}{g n_\chi(t)}\right)^{1/(n+1)}.
\end{equation}
The above equation suggests that the mass of the preheat field does
depend on the cosmological time through the evolution of $n_\chi$.
From these equations we can calculate the ratio of $\rho_\chi$
to $V(\sigma)$
\begin{equation}
\frac{\rho_{\chi}}{V(\sigma)}=n.
\end{equation}
It is now clear that the ratio is given by the constant $n$.
Note that the energy density of the preheat field is always larger than
the curvaton potential energy.
Since the number density $n_\chi$ evolves as $n_\chi\propto
a^{-3}$, the energy density of the preheat field $\rho_{\chi}$ will
evolve as
\begin{equation}
\rho_{\chi}\propto a^{-3(1-\frac{1}{n+1})}
\end{equation}
keeping the ratio $\frac{\rho_{\chi}}{V(\sigma)}=n$.
Therefore, the evolution of the preheat field density is slightly slower
than the standard matter $\rho_m \propto a^{-3}$. 
Note that the mass of the preheat field $m_\chi$ grows as
\begin{equation}
m_\chi \simeq g \sigma \propto a^{\frac{3}{n+1}}.
\end{equation}

The decay of the preheat field is determined by its couplings to light
fermions $\sim g_\chi \chi \bar{\psi} \psi$, where
the preheat field is supposed to decay at $H\simeq \Gamma_\chi$.
If the decay rate is  given by
\begin{equation}
\Gamma_\chi \simeq m_\chi^3/M_p^2,
\end{equation}
the preheat field will decay when $\sigma$ reaches the value
\begin{equation}
\sigma_d \simeq \left(\frac{M_p M^{\frac{n+4}{2}}}{g^3}
\right)^{\frac{2}{n+6}}.
\end{equation}
Here we have assumed that the Universe is already dominated by the preheat
field when it decays (i.e. the Hubble constant is given by $H^2 \simeq
\rho_\chi/M_p^2\simeq V(\sigma)M_p^2$). 
Suppose that the inflaton energy density is initially $\rho_I \simeq
v^2$ and then evolves\footnote{See for examle
Ref.\cite{dimo_trapped} for the evolution of the inflaton energy
density during the trapped oscillation.}
 as $\rho_I \propto a^{-4}$, and also that the preheat field is   
initially $n_\chi^i \simeq v^{3/2}$ and then evolves as
$n_{\chi}\propto a^{-3}$.
Here the condition for the curvaton domination is $\rho_I^d \le
g\sigma_d n_\chi^d $. 
The preheat-field number density at the decay is determined by the
value of $\sigma_d$ and the relation (\ref{rela_si}), and  becomes
$n_{\chi}^d \simeq \frac{M^{n+4}}{\sigma_d^{n+1}}$.
Therefore, the inflaton density at that time is 
\begin{equation}
\rho_I^d \simeq v^2 \left(\frac{n^d_\chi}{n_\chi^i}\right)^{4/3}
\simeq M^4 \left(\frac{M^{n+1}}{\sigma_d^{n+1}}\right)^{4/3}.
\end{equation}
For example, considering $n=4$ and $M=10^{-16}M_p$ we obtained
$\sigma_d \simeq 10^{3} M$, which suggests that
 $\rho_I^d/\rho_\chi^d \sim 10^{-20}$.
Obviously, it is very easy to satisfy the required condition for
curvaton (preheat-field) domination. 
If there is a cut-off for the effective potential $V(\sigma)$,
the curvaton $\sigma$ will be trapped at the origin with (perhaps)
a constant value of the potential until the preheat
field is diluted below the critical value.
Then the evolution is determined by the number density of the preheat
field, which does not induce any difference to the above calculation.

After the decay of the preheat field, the curvaton $\sigma$ does not
experience the confining potential that had been induced by the preheat
field. 
The effective mass of the curvaton at that time
is $m_\sigma \simeq M^{n+4}/\sigma_d^{n+2}$.
Considering again the case with $n=4$ and $M=10^{-16}M_p$, we obtain
the ratio between the curvaton mass and the Hubble constant just after
the decay; $m_\sigma^2 /H^2 \sim 10^{26}$.
This result suggests that the curvaton starts to roll down the potential
and rapidly approaches the tracker solution. 
It is always possible to introduce fine-tuning of the cosmological
constant so that our model fits the cosmological observations. 
One may also consider various kinds of quintessential potential for the
NO curvatons, which may or may not explain the dark energy of the
Universe depending on the later quintessential scenario. 
In any case, the fine-tuning of the cosmological constant will remain
a serious problem until we can discover the secret related to the
radiative corrections, the symmetry breaking and the origin of the
cosmological constant. 

\section{Conclusions and Discussions}
\hspace*{\parindent}
\label{sec:conclusion}
We considered the curvaton scenario in which the curvaton has the
non-oscillatory potential. 
The primordial isocurvature perturbation of the NO curvaton is converted
into the number density of the preheat field when the preheat field is
generated through preheating. 
The evolution of the curvaton and the preheat field is controlled by the
trapping mechanism until the preheat field decays to reheat the
Universe. 
Since the densities of the NO curvatons (and the preheat field) evolve
slightly slower than the standard matter, the ratio between these fields
and the total energy density grows with time. 
At the same time, the mass of the preheat field grows as the NO curvaton
rolls down the potential. 
A peculiar feature of the scenario is that the late-time domination and
the reheating is induced by the preheat field, not by the NO curvaton
itself. 
Our basic idea is (1) the nucleation rate of the cosmological relic
density is fluctuated by the primordial isocurvature perturbations, (2)
the rate of the massive relic density to the total energy density of the
Universe grows during the cosmological evolution, and (3) eventually
the massive relic decays to reheat the Universe leaving behind the
curvature perturbation. 
In our present model the light field ("NO curvatons") accompanied by the
primordial isocurvature perturbation is responsible 
for neither the domination nor the reheating. 
This peculiar feature of the NO curvaton is of course very different
from the conventional curvaton scenario. 
In our scenario the isocurvature perturbation of the curvaton is used
solely to bias the nucleation rate of the preheat field. 
The preheat field (not the curvaton itself) is responsible for the
late-time domination and the reheating. 
The situation is very similar to the "alternatives" to the traditional
inflationary scenario, in which the role of the traditional inflaton is
shared with the additional light field, making the constraints on the
inflaton potential looser than the traditional ones.
The NO curvaton shares the role of the traditional curvatons with the
preheat field, making the constraints on the curvaton potential very
different from the traditional curvaton scenario. 

Moreover, NO curvatons could be considered as the hybrid version of the
quintessential inflation model. 
In this hybrid model one does not have to look for a potential that has
the property of chaotic-type inflaton in the left-hand side while it has
run-away behavior in the right-hand side. 
Finding such a peculiar potential has been a serious obstacle for the
single-field scenario. 
In the hybrid model the conventional inflaton potential and the
quintessential potential can be separated and are connected at the ESP. 

\section{Acknowledgment}
We wish to thank K.Shima for encouragement, and our colleagues at
Tokyo University for their kind hospitality.

\end{document}